\pdfoutput=1

\documentclass[sigconf]{acmart}

\copyrightyear{2026}
\acmYear{2026}
\setcopyright{cc}
\setcctype{by}
\acmConference[UIST Adjunct '26]{The 39th Annual ACM Symposium on User Interface Software and Technology}{November 02--05, 2026}{Detroit, MI, USA}
\acmBooktitle{The 39th Annual ACM Symposium on User Interface Software and Technology (UIST Adjunct '26), November 02--05, 2026, Detroit, MI, USA}
\acmDOI{10.1145/3830397.3841870}
\acmISBN{979-8-4007-2855-6/2026/11}

\usepackage{balance}
\usepackage{microtype}
\begin{document}

\title{Feelium: A Touchable Blimp Body for Aerial Telepresence}

\author{George Xi Wang}
\affiliation{%
  \institution{Stony Brook University}
  \city{Stony Brook}
  \state{New York}
  \country{United States}}
\affiliation{%
  \institution{New York University}
  \city{Brooklyn}
  \state{New York}
  \country{United States}}
\email{george.x.wang@stonybrook.edu}
\email{xw3617@nyu.edu}

\author{Henghao Li}
\affiliation{%
  \department{School of Science and Engineering}
  \institution{The Chinese University of Hong Kong, Shenzhen}
  \city{Shenzhen}
  \state{Guangdong}
  \country{China}}
\email{lihenghao@cuhk.edu.cn}

\author{Shan Lin}
\affiliation{%
  \institution{The Chinese University of Hong Kong, Shenzhen}
  \city{Shenzhen}
  \state{Guangdong}
  \country{China}}
\email{123090334@link.cuhk.edu.cn}

\author{Yunge Wen}
\affiliation{%
  \department{MIT Media Lab}
  \institution{Massachusetts Institute of Technology}
  \city{Cambridge}
  \state{Massachusetts}
  \country{United States}}
\email{yungew@mit.edu}

\author{Jiaqian Hu}
\affiliation{%
  \department{Translation and Localization Management}
  \institution{Middlebury Institute of International Studies at Monterey}
  \city{Monterey}
  \state{California}
  \country{United States}}
\email{jiaqianh@middlebury.edu}

\author{Yuhua Jin}
\affiliation{%
  \department{School of Science and Engineering}
  \institution{The Chinese University of Hong Kong, Shenzhen}
  \city{Shenzhen}
  \state{Guangdong}
  \country{China}}
\email{yuhuajin@cuhk.edu.cn}

\renewcommand{\shortauthors}{Wang et al.}

\begin{teaserfigure}
  \centering
  \includegraphics[width=\textwidth]{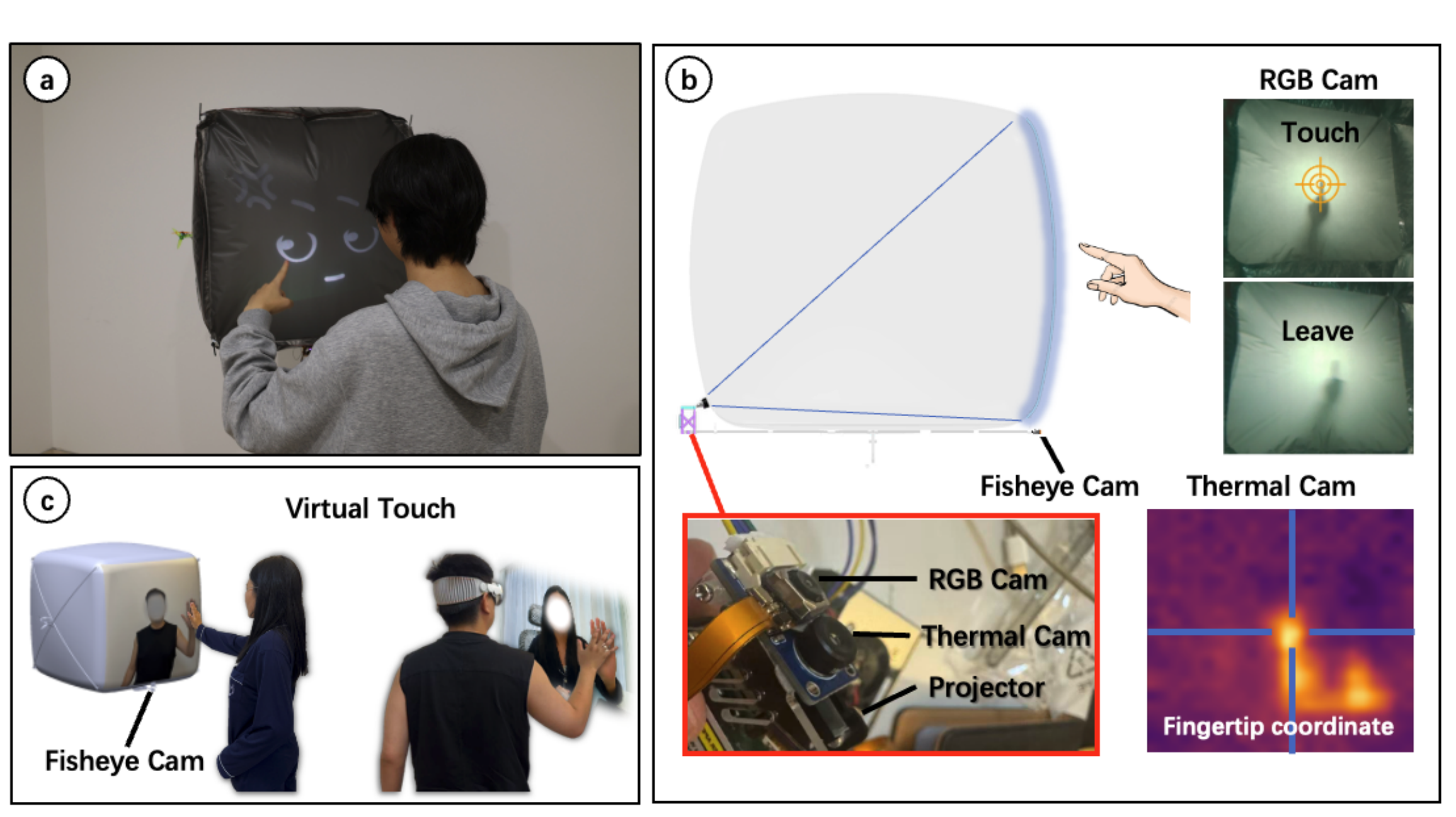}
  \caption{\emph{Left:} the remote person inhabits the blimp through a VR
  headset and meets a partner's touch first-person. \emph{Right:} two cameras
  split the roles: a front-facing fisheye camera gives the wearer wide-angle
  eyes, keeping a reaching hand in view until it lands, while a rear RGB
  camera (with a thermal companion) senses that touch as a shadow on the
  back-lit skin.}
  \Description{Left: two stacked panels labelled Virtual Touch and VR headset
  show a partner pressing a palm on a cube blimp's projected face, and a
  person in a VR headset whose avatar appears on the floating blimp. Right:
  the blimp with an inset photograph of the onboard module labelling an RGB
  camera, a thermal camera, and a projector; two camera frames labelled Touch
  and Leave show a fingertip shadow on the membrane; a thermal frame marks a
  warm fingertip coordinate.}
  \label{fig:teaser}
\end{teaserfigure}

\begin{abstract}
Floating things invite touch. We present \textbf{Feelium}, a blimp-based telepresence platform that enables visual embodiment and touch interaction through its inflatable skin. Through a VR headset, a remote person inhabits the blimp, looking out of it first-person, appearing on its skin as a face or avatar, and steering it through the room. Partners in the room pat it, press a palm against it, draw on it, or lean into it; the skin senses each contact, renders it into the wearer's view in VR spaces.Touch thus provides a physical interaction channel for remote presence, turning the skin into a shared surface between remote and co-located partners.

\end{abstract}



\maketitle

\section{Introduction}

People reach out to touch what floats beside them: when Abtahi et al.\ wrapped a
quadcopter in a safe-to-touch cage, touch leapt from 1.4\% to 38.9\% of
interactions~\cite{abtahi2017dronenearme}. However, the system cannot pass the
touch back to the person operating the drone. Later haptic and touchable drones kept
the same limits: contact stays fingertip-scale and transient, sensing lives
in external tracking systems, and the rotors stay loud and
dangerous~\cite{abtahi2019beyondtheforce, hoppe2018vrhapticdrones,
abdullah2017hapticdrone, nitta2014hoverball, gomes2016bitdrones}.

Blimps are the opposite. They can hover among people for hours and be
touched without risk. Two decades of
research have explored blimps as carriers and screens: tele-embodiment
(PRoP~\cite{paulos1998prop}), projected
faces~\cite{tobita2011floatingavatar}, flying
signage~\cite{tobita2014aeroscreen}, light-field heads
(LightBee~\cite{zhang2019lightbee}), ultrasound-positioned midair
balloons~\cite{furumoto2021midairballoon}, and flapping-wing companions
built for closeness rather than input~\cite{xu2025spreadyourwings,
xu2025cuddlefish}. Across these systems the envelope is output-only.
Touchable inflatables do exist on the ground --- hemispherical multi-touch
displays~\cite{stevenson2011inflatable}, Emoballoon's unlocalized social
touch~\cite{nakajima2013emoballoon}, robot-reeled tangible
balloons~\cite{pham2025buoyance} --- yet \emph{none connects the touched
surface to a remote person}. Touch, though, is precisely the channel that
deepens mediated closeness, from remote
handshakes~\cite{nakanishi2014handshake} to cortisol-lowering huggable
media and multi-task VR space~\cite{sumioka2013hugvie, qian2025duozone}.

\textbf{Feelium} fills this gap: to our knowledge, it is the first to explore envelope touch in an aerial telepresence context. We contribute: (1) the Feelium system, which integrates a flying-display blimp platform, first-person VR inhabitation, and an onboard touch-sensing envelope; (2) a vocabulary of two-party touch interactions between a co-located partner and the remote person inside; and (3) a preliminary technical validation of touch sensing on an inflatable envelope.

\section{The Feelium System}

\textbf{Blimp.} Feelium's body is a 64\,cm indoor helium blimp. Most of
the envelope is blackened to retain helium and reject ambient light; the
light-scattering front face is a rear-projection screen, lit from inside by
a 24.9\,g laser projector (Ultimems HD309) through a 180$^\circ$ fisheye
lens. Four motors under closed-loop
control hold a quiet, stable hover among people, and the complete body
weighs 323.7\,g against ${\sim}360$\,g of buoyant lift.

\textbf{VR.} A remote person puts on an Apple Vision Pro
and \emph{becomes} the blimp (Figure~\ref{fig:teaser}, left).
A front-facing fisheye camera serves as the wearer's eyes; its wide field
keeps a partner's reaching hand in view all the way to contact, so a touch
arrives on screen rather than out of frame. The wearer's face or animated avatar is displayed
on the envelope so partners see who is inside.
Six motion commands (forward, backward, up, down, turn left, turn right)
let the wearer
drift, turn, and station the body in the room; control, video, and speech all
travel over HTTP.

\textbf{Touch.} A press from
outside casts a shadow in the projector's back-light. A second camera at the
rear of the blimp (a Raspberry Pi AI Camera, Sony IMX500)
watches the membrane from inside, paired with a 32$\times$24
MLX90640-D55 thermal camera (Figure~\ref{fig:teaser}, right). The two split
the roles cleanly: the RGB camera makes every touch decision, while the
thermal camera senses only presence. The RGB shadow
pipeline~\cite{matsushita1997holowall} keeps a slowly adapting reference
image of the untouched, back-lit
membrane and computes per-pixel \emph{relative shadow depth}: the fraction of
back-light lost against that reference, which makes the signal independent of
the projector's uneven brightness. A shadow counts as touch when it passes
either of two gates. An instant depth gate catches firm presses. A
\emph{sustained-anchoring} gate takes the rolling minimum of shadow depth over the
past second: a pressing fingertip stays pinned to the membrane,
while a hovering hand wobbles and its shadow momentarily thins. The fingertip is localized as
the centroid of the shadow's darkening core. A tracker then segments the
per-frame decisions into discrete events (onset to release, with debouncing),
distinguishing \emph{tap}, \emph{hold}, and \emph{swipe} by contact duration
and fingertip path length. The thermal camera complements this. Each frame is
differenced against a slowly adapting ambient baseline, and a cluster of
pixels a few degrees above it reads as a warm body, letting the blimp notice
a hand or partner drawing near before and between contacts. Thermal frames
carry no touch decision (conducted heat through the membrane is too weak and
slow). Each detected touch is rendered as a located
marker in the wearer's first-person view, and the projector answers at the
touched spot.

\section{Interacting with an Inhabited Blimp}

The skin's contact primitives, composed over time and area, turn everyday
gestures into conversation between the partner in the room and the person
inside. \textbf{Greeting pat}: taps on the ``head'' ripple into the wearer's
view; the envelope returns a warm pulse. \textbf{Palm-to-palm}: the wearer
raises a hand to meet a palm-shaped glow, two prints merging --- held hands
without hands. \textbf{Peck tap}: a tap at the ``cheek'' blooms a small
heart; the wearer can pinch to send one back. \textbf{Finger-drawing guess}: strokes
render live in the wearer's view, and a correct guess ignites the drawing.
\textbf{Lean-in hug}: sustained large-area contact sets the whole skin
glowing in a slow breathing rhythm. \textbf{Wearer-initiated}: a projected
shimmer invites the partner's hand, and small motion replies (a drift
closer, a gentle bounce) let the body respond. These exchanges suit settings
where presence, not information, is the payload: a goodnight palm-press in a
couple's daily call, a drawing game with a child before bed, or a distant
relative drifting through a family gathering as a body people can pat hello.

\section{Preliminary Results}

\begin{figure}
  \centering
  \includegraphics[width=\columnwidth, trim=0 90pt 0 10pt, clip]{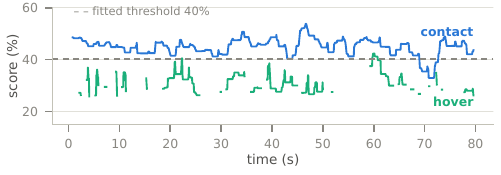}
  \caption{Sustained-anchoring score on labelled contact and hover clips:
  contact stays almost entirely above the fitted threshold; hover keeps
  dipping below.}
  \Description{A time-series plot in which the contact clip's score stays
  almost entirely above a dashed threshold line while the hover clip's score
  repeatedly dips below it.}
  \label{fig:results}
\end{figure}

We validated the skin on recordings from the real blimp
(Figure~\ref{fig:results}): 228 labelled trials over three sessions --- 138 prompted touches (78 taps, 24 holds, 36 swipes) and 90
negatives (30 hovers, 30 no-contact hand disturbances, 30 empty scenes) ---
plus three continuous ${\sim}75$\,s clips (steady contact, hovering, empty).
\emph{Detection:} the shadow detector is sensitive: with a hand present
(touching or not), it registered activity in 171 of 198 trials, and it never
fired in the 30 empty scenes. The hard decision is therefore not seeing the
hand but \emph{rejecting near-hovers}: thresholding the darkened area alone
reaches 95\% recall but only 77\% precision, because a hand hovering just
above the membrane darkens it almost as much as a touch. The
sustained-anchoring gate closes this gap: on the continuous clips its score
separates contact from hover at AUC~0.99, and a threshold fit on the first
60\% of each clip holds 0.90 balanced accuracy on the unseen
remainder. \emph{Localization:} the 69 taps and holds with a detected
landing point cluster tightly by prompted target (RMS spread 4--21\,px,
median 9\,px in the detector's 640$\times$480 working resolution);
coordinates are camera-space, as
screen-to-camera calibration is future work. The pipeline runs onboard at
${\sim}15$\,fps. Current limits: results are
from dim-room lighting and one rectangular ``sensitive skin patch'' of the
envelope.

\section{Conclusion and Future Work}

Feelium enables touchable, inhabitable aerial telepresence, integrating an
onboard touch-sensing envelope with first-person VR embodiment for two-party
social touch. A preliminary technical validation provided early evidence of
the system's feasibility and potential, motivating a dyadic evaluation of
social presence with the full touch loop as well as whole-envelope sensing,
haptic feedback to the wearer, and gesture-piloted flight. Running fully
onboard a quiet indoor blimp, this work hopes to give remote presence a body
that invites touch, bringing separated partners and families within arm's
reach.

\begin{acks}
This work was supported by the University Development Fund of The Chinese
University of Hong Kong, Shenzhen under Grant UDF01003513.
\end{acks}

\balance
\bibliographystyle{ACM-Reference-Format}
\bibliography{references}

\end{document}